\documentclass[sigconf]{acmart}
\renewcommand\footnotetextcopyrightpermission[1]{} 
\usepackage{tabularx}
\usepackage[utf8]{inputenc}
\usepackage{breakurl}
\usepackage{longtable}
\usepackage{multirow}
\usepackage{xspace}
\usepackage{footmisc}

\newcommand{\update}[1     ]{{\color{magenta} #1}}

\newcommand{\oana}[1]{{\emph{\color{red}#1 --\textbf{Oana}}}}

\newcommand{\adanalystbr}{\textsc{AdCollector}\xspace}

\newcommand{\Fbdata}{\textsc{FbAdLibrary}\xspace}
\newcommand{\gs}{\textsc{Gold Standard}\xspace}
\newcommand{\adanalyst}{AdAnalyst\xspace}
\newcommand{\admonitor}{Facebook Ad Monitor}
\newcommand{\polads}{\textsc{Political Ads}\xspace}
\let\ACMmaketitle=\maketitle
\renewcommand{\maketitle}{\begingroup\let\footnote=\footnote \ACMmaketitle\endgroup}
\AtBeginDocument{%
  \providecommand\BibTeX{{%
    \normalfont B\kern-0.5em{\scshape i\kern-0.25em b}\kern-0.8em\TeX}}}

\acmYear{2020}
\acmConference[WWW '20]{Proceedings of The Web Conference 2020}{April 20--24, 2020}{Taipei, Taiwan}
\acmBooktitle{Proceedings of The Web Conference 2020 (WWW '20), April 20--24, 2020, Taipei, Taiwan}
\acmPrice{}

\makeatletter
\def\@copyrightspace{\relax}
\makeatother
\begin{document}

\title{Facebook Ads Monitor: An Independent Auditing System for Political Ads on Facebook}
\titlenote{This is a preprint version of a paper that will appear at WWW’20 Conference.}
\author{Márcio Silva}

\affiliation{%
  \institution{Univ. Federal de Mato Grosso do Sul}
  \department{Faculty of Computer Science}
  \city{Campo Grande}
  \state{}
  \country{Brazil}
}

\author{Lucas Santos de Oliveira}
\affiliation{
  \institution{Univ. Estadual do Sudoeste da Bahia}
  \department{Science and Technology Department}
  \city{Jequié}
    \state{}
  \country{Brazil}}

\author{Athanasios Andreou}
\affiliation{%
  \institution{Univ. Grenoble Alpes, CNRS, Grenoble INP, LIG}
  \city{Grenoble}
  \state{}
  \country{France}
}

\author{Pedro Olmo Vaz de Melo}
\affiliation{%
 \institution{Universidade Federal de Minas Gerais}
  \department{Computer Science Department}
 \city{Belo Horizonte}
 \state{}
 \country{Brazil}
 }

\author{Oana Goga}
\affiliation{%
  \institution{Univ. Grenoble Alpes, CNRS, Grenoble INP, LIG}
  \city{Grenoble}
  \state{}
  \country{France}
  }

\author{Fabrício Benevenuto}
\affiliation{%
  \institution{Universidade Federal de Minas Gerais}
  \department{Computer Science Department}
  \city{Belo Horizonte}
  \state{}
  \country{Brazil}
  \postcode{78229}}



\renewcommand{\shortauthors}{}

\begin{abstract}
  The 2016 United States presidential election was marked by the abuse of targeted advertising on Facebook. Concerned with the risk of the same kind of abuse to happen in the 2018 Brazilian elections, we designed and deployed an independent auditing system to monitor political ads on Facebook in Brazil. To do that we first adapted a browser plugin to gather ads from the timeline of volunteers using Facebook. We managed to convince more than 2000 volunteers to help our project and install our tool. Then, we use a Convolution Neural Network (CNN) to detect political Facebook ads using word embeddings. To evaluate our approach, we manually label a data collection of 10k ads as political or non-political and then we provide an in-depth evaluation of proposed approach for identifying political ads by comparing it with classic supervised machine learning methods.
Finally, we deployed a real system that shows the ads identified as related to politics. 
We noticed that not all political ads we detected were present in the Facebook Ad Library for political ads.  
Our results emphasize the importance of enforcement mechanisms for declaring political ads and the need for independent auditing platforms. 
\end{abstract}




\maketitle

\section{Introduction}




%



The 2016 United States presidential election was marked by an information war that took place into different social media platforms~\cite{badawy2018analyzing,lima2018@asonam}. In particular, the election was marked by the abuse of targeted advertising on Facebook;
a group of Russian citizens and companies were indicted by U.S. authorities for trying to influence the 2016 US election through the Facebook Ad Platform~\cite{russian-ads,cnnrussian}. A recent study characterized a set of ads released by the house of representatives posted by the Russian company named Internet Research Agency (IRA)~\cite{ribeiro2019@fat}. Their findings show that these ads received a click-through rate about 10 times higher than the one for typical ads in Facebook, that ads were ran along two years and not only during the 2016 election, and that the content of these ads was on polarizing topics (e.g., immigration, race-based policing), usually targeting vulnerable sub-populations. 
The case of the IRA ads clearly raised numerous concerns about external interference in elections through targeting advertising.  Beyond that, it showed how an ad platform could be used for posting illicit political ads targeting people susceptible to false stories, stoke grievances, and incite social conflict. In other words, a platform of this kind could be abused to engineer polarization in a country, which could end up favoring a political campaign. 


As an answer to these new threats, Facebook implemented several countermeasures. On May 24, 2018, Facebook changed its ToS policy to allow the launching of political ads only by advertisers that reside in the same country with the people targeted (this requirement does not apply to non-political ads)~\cite{tos_change_facebook}. Also, all election-related ads must be clearly labeled as such, including a “Paid for by” disclosure from the advertiser at the top of the ad.
On June 28, 2018, Facebook launched a page to show the ``active campaigns'' of advertisers that send political ads that are accessible from their corresponding Facebook Pages~\cite{facebook_adarchive_lached} (this service was a test pilot only available in Canada before the Brazilian elections between Aug. 16 -- Oct. 28, 2018). 
Finally, in June 2018 Facebook launched the Facebook Ad Library~\cite{facebook_ad_archive_launched, facebook_ad_archive}, a service that allows people to see all the ads declared as \emph{political} by the  advertisers posting them  as well as information about who paid for the ad, the amount spent, the number of impressions delivered, and the audience. 
A few weeks after that, Facebook launched the Ad Library to Brazilian citizens~\cite{facebook_ad_archive_launched}. 



While these measures were welcomed, many people including researchers, journalists and organisations pointed out that they are not sufficient~\cite{efective_ad_archive, mozillaadlibraryreport, cnnAdLibraryDontWork}.
\textit{First}, advertisers have to declare themselves, on a voluntary basis, whether they are sending political ads. 
This is problematic because dishonest political parties and presidential candidates can avoid scrutiny of their ad messages by not declaring them as political. \textit{Second}, beyond public opinion manipulation and spread of fake news, the Facebook ads platform can also be used for \emph{slush funds}~\cite{fabriciofolha}. Brazilian electoral law states that companies are prohibited to make donations to any political party or candidate during the election period. Currently, dishonest companies can spend an unlimited amount of undeclared money in favor of a political agenda through the Facebook ads platform~\cite{electoral_law}. 
\textit{Third}, the Facebook Ad Library does not offer information regarding the targeting used by advertisers. A recent study found out that Facebook makes available to advertisers more than 240,000 interests that it has inferred about its users, such as ``Yoga'', ``Gluten-free'', or ``Adult Children of Alcoholics''~\cite{speicher-2018-targeted} . Combinations of these interests can result in uncontrollable ways of reaching very specific sub-populations vulnerable to specific messages. 

\begin{figure}[t]
\includegraphics[width=1.0\columnwidth]{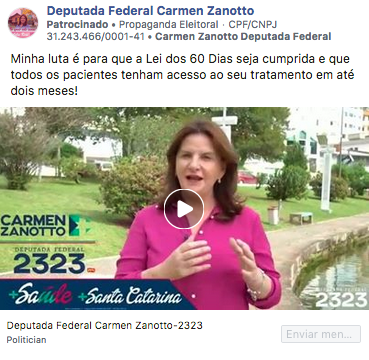}
\caption{Example of an official political ad posted during the election period in Brazil.}
\label{fig:ad_anastasia}
\end{figure}

Advertisers of political content are supposed to both comply with the Facebook's Terms of Service as well as the election legislation in their respective countries.  
In October 2017, Brazilian authorities demanded that political figures that are advertising on Facebook political content along the electoral period, an established period near the elections, need to give information about their national identification numbers, namely CPF, for individuals, and CNPJ, for companies. Facebook responded by creating an interface that allows advertisers of political content to include disclosure information in their ads related to elections and also their CNPJs or CPFs (see Figure~\ref{fig:ad_anastasia}). However, Facebook we are unsure whether they deployed any enforcement mechanisms for tagging political ads that try to run without the right disclaimer in Brazil.

Concerned by the eminent high potential misuse of Facebook ads and imminent risks to Brazilian electoral laws, we designed and deployed a system to monitor political ads on Facebook, named \admonitor\footnote{\url{https://www.eleicoessemfake.dcc.ufmg.br/anuncios}}\footnote{The extension has been forked from \url{adanalyst.mpi-sws.org}.}. 
Our tool is a Chrome and Firefox extension that users can install on their computers and that collects the ads users see when they check their Facebook timeline as well as the corresponding explanations from the ``Why am I seeing this ad?'' feature that reveals some information about the targeting used by the advertisers~\cite{Andreou19a}. Our tool is similar with the tools provided by ProPublica~\cite{propublicahome} and WhoTargetsMe~\cite{whotargetsme}, but it is customized for collecting and analyzing ads in Portuguese besides English.
In addition, we developed a web application that runs our political ad classifier and allows Brazilian authorities and citizens to monitor the Facebook ads that our browser extension has collected. The classifier calculates the political probability score for each ad we collect. Our web application has a search engine where anyone can search and perform filters over our data.  Our hope was that the detection of a single illegal ad is enough to incur strict penalties and, hopefully, to inhibit the proliferation of such ads. 




To disseminate the tool we wrote and published opinion articles on the media disclosing ways  in which online systems can be exploited to influence elections and what we can do about them\cite{fabriciofolha, nyt2018benevenuto}.  We also presented these threats in the Brazilian senate.\footnote{\url{https://www12.senado.leg.br/noticias/materias/2018/05/11/impacto-das-midias-sociais-para-o-legislativo-sera-discutido-em-seminario-no-senado}}\footnote{\url{https://www.youtube.com/watch?v=eGScrdi5hhU&t=3450s}} 
Our tool was installed by more than 2000 users, out of which 715 users actively used the tool along the election period, providing us all the ads they received while navigating on Facebook. This collaborative effort provided us with a dataset containing 239k ads from 40k advertisers along the period of March 14, 2018 to October 28, 2018.


We implemented several machine learning-based techniques for detecting political ads on Facebook. 
We tested supervised classifiers such as Naive Bayes, Random Forest, Logistic Regression, SVM and Gradient Boosting as well as a recently proposed Convolution Neural Network (CNN) built to detect political tweets.
 To evaluate our algorithms we created a golden standard test collection with 20k ads labeled as political or non-political.\footnote{\label{dataset}Dataset and code available at \url{https://lig-membres.imag.fr/gogao/political_ads.html}.} Our results show that the CNN-based model is able to achieve an AUC of 98\% and  an accuracy of 94\% in a near balanced dataset or a 78\% true positive rate for a 1\% false positive rate.

We tested the CNN classifier over a dataset of 38k ads containing ads in Portuguese during the electoral period (August 16, 2018 to October 28, 2018), and we found 835 (approx 2\%) of the ads to be political. Even if we only have a small sample of ads running on the platform, our results show that there are many ads with political content that were detected by our algorithm and were not labeled as such on Facebook. 

Our study emphasize the importance of enforcement mechanisms for declaring political ads.  The big open questions, however, are (i) who is responsible for enforcing the election legislation?; and (ii) how can authorities be able to enforce laws if they do not have access to data?
While companies and politicians are debating how to regulate political ads \cite{twitter_ban_political_ad, microtargeting_europe}, we believe one potential solution could be independent auditing platforms such as ours that collect ads from volunteers and search for undeclared political ads. We show that it is feasible to build such a platform and have a positive impact in the real world.   Our auditing platform is online (\url{https://adanalyst.mpi-sws.org/}) and we plan to monitor other elections as well.  Furthermore, to facilitate the development of new auditing platforms, our code is open source and some parts of our data are public.\footref{dataset}

The rest of the paper is organized as follows. Section~\ref{political_ads} provides the necessary background on the Brazilian election legislation and how the Facebook ad platform works. Section~\ref{sec:dataset} describes our dataset and data collection method. Section~\ref{sec:methodology} describes our algorithms for detecting political ads and Section~\ref{sec:results} describes the results and findings after deploying our system in the real-world. Finally, Section~\ref{related-work}, highlights our contributions in comparison with related studies. 



\section{Background} \label{political_ads}

This section provides background on the Brazilian election and legal requirements for advertising political content, as well as how advertising on Facebook works, what are the platform's Terms of Service regarding political content, and transparency mechanisms provided. 
Note that there is a distinction between the legal requirements from state legislation that advertisers have to follow, and the requirements advertisers need to respect due to the ToS of Facebook.

\subsection{Brazilian Election Legislation} 

In Brazil elections occur every four years and are divided between national and local elections. In national elections, Brazilian citizens choose their president, governors, deputy-governors, state representatives, senators and congressmen in all 26 states and the Federal District. Local elections are for choosing mayors and councilors. 


The 2018 presidential elections in Brazil had two rounds. To win the election directly in the first round a candidate should have at least 50\% of the valid votes, which was not the case in 2018. The voting period for both rounds last only a single day. In 2018, the first round was in October 7 and the second round in October 28. From August 16 onward, \emph{electoral advertising}, such as rallies, caravans, distribution of graphic material and advertising on the Internet was permitted.




For electoral advertisements on the Internet, Resolution No. 23,551 of the Superior Electoral Court of Brazil~\footnote{\url{http://www.justicaeleitoral.jus.br/arquivos/resolucao-23551-nova/rybena_pdf?file=http://www.justicaeleitoral.jus.br/arquivos/resolucao-23551-nova/at_download/file}} stipulated a series of rules in order to ensure greater transparency for campaign spending. For advertisements on social networks, such as on Facebook, chapter IV of the resolution stipulates that they can be made or edited by:
\begin{itemize}
    \item candidates, political parties or coalitions;
    \item any natural person (e.g. a bot or a fake profile \emph{is not} a natural person), \textit{as long as they do not pay for content promotion}. Meaning that they can only promote candidates through regular posts and not sponsored ads. 
\end{itemize}
In addition, any promoted content must contain, in a clear and legible form, the registration number in the National Register of Legal Entities (CNPJ)~\footnote{An identification number issued to Brazilian companies by the Department of Federal Revenue of Brazil.} or the registration number in the Register of Individuals (CPF)~\footnote{The Brazilian individual taxpayer registry identification.} of the responsible person, in addition to the expression \emph{``Electoral Advertising''} (in Portuguese: \emph{Propaganda Eleitoral}).

Also, Paragraph 4 of Article 23 of the resolution stipulates that the Internet application provider (e.g., Facebook) that allows paid political content to be promoted must clearly communicate this to its users. In compliance with this law, Facebook provides an interface for political agents to properly register their advertisements during the election period. Figure~\ref{fig:ad_anastasia} shows an example of a political ad properly promoted on Facebook using this interface. All the respective expenses with those ads needed to be declared to the Superior Electoral Court.


%

\subsection{Regular Advertising on Facebook}

Any user with an active profile and Facebook page can become an advertiser. All they need to do is activate their ad account, select a targeting audience, fine-tune parameters such as bidding, and they can automatically send ads to their desired audience~\cite{tospolitical}. This means that everyone with a Facebook account can spread content about products, ideas, or even malicious information and fake news.

Facebook offers a multitude of audience selection options that can enable advertisers to target in a way which is considerably more fine-grained than traditional online advertising platforms like Google~\cite{google}. Apart from traditional targeting options, like \emph{age}, \emph{gender}, \emph{location}, and \emph{language}, advertisers can use - and combine in formulas - a variety of \emph{attributes} to target users. 
Advertisers can target users that have their birthday next month, or are interested in common subjects like \emph{Games} or \emph{Food}, but they can also target users interested in much more sensitive attributes such as \emph{Homosexuality}, or \emph{Fascism}. In fact, there exist more than 240,000 available attributes for advertisers to choose from~\cite{speicher-2018-targeted}, including at least 2092 potentially sensitive attributes~\cite{cabanas2018unveiling}. Additionally, advertisers can target users through \emph{custom audiences}~\cite{speicher-2018-targeted} by uploading to Facebook lists of users' Personally Identifiable Information (PII), such as phone numbers, emails, or names and physical addresses. Or, they can target \emph{lookalike audiences}, users that resemble some another desired user group according to Facebook~\cite{lookalike}.  These targeting options are naturally not only at the disposal of benign advertisers, but also malicious advertisers with ill intents.

\subsection{Political Advertising on Facebook}

Regardless of Brazilian (or any other country) regulations, advertisers who want to send political ads on Facebook are subject to higher levels of scrutiny and the corresponding ads are subject to higher levels of transparency. Facebook defines an ad as political when~\cite{facebook_issue_content, facebook_political_content}:
\begin{trivlist}
\item \textit{(i) it is made by, on behalf of or about a current or former candidate for public office, a political party, a political action committee or advocates for the outcome of an election to public office; or}
\item  \textit{(ii) it relates to any election, referendum or ballot initiative, including "get out the vote" or election information campaigns; or (iii) it relates to any national legislative issue of public importance in any place where the ad is being run; or}
\item \textit{(iv) it is regulated as political advertising.}
\item \textit{(v) it is related to issues of public importance: abortion, budget, civil rights, crime, economy, education, energy, environment, foreign policy, government reform, guns, health, immigration, infrastructure, military, poverty, social security, taxes, and terrorism}
\end{trivlist}
Note that this is not the only definition for what is a political ad and scholars and regulators are currently debating about definitions~\cite{whatispoliticalfacebook}.

To sponsor political content on Facebook, advertisers need to first verify their accounts, declare that their ad is about political or social issues and put a disclaimer that mentions who paid for the ad.\footnote{\url{https://www.facebook.com/business/help/208949576550051?id=288762101909005}}

\subsection{The Facebook Ad Library}
The ads declared by  advertisers as containing political content are part of the historical Facebook Ad Library. 
While Facebook provides functionalities for advertisers to include disclosure information about their political ads, we do not know whether Facebook has any kind of enforcement mechanisms for detecting political ads that do not have the appropriate disclosure in Brazil.

\section{System for Monitoring Ads}
\label{sec:dataset}

This section describes our tool and the datasets we collected for this study.

\subsection{Tool Design}

Our tool is a browser extension (for Chrome and Firefox) that collects the ads that appear in the Facebook feed of the volunteers who install it. 
We forked \adanalystbr from \adanalyst\footnote{\url{https://adanalyst.mpi-sws.org/}}, and we added support for collecting and analyzing ads in the Portuguese language as well as an interface in Portuguese where users can browse the ads we collect. 



To capture the ads that users receive on Facebook, we scrape the Facebook's HTML and we look  for the tag \emph{``Sponsored''} (\emph{``Patrocinado''} in Portuguese). This tag is used by Facebook to help users distinguish sponsored content from the rest. 
The captured frame contains the media content of the ad (either a video, an image, or a collection of images), the text of the ad, and a link to the advertiser's page. Our browser extension does not collect ads that appear when a user is watching a video on Facebook.

Facebook provides explanations to users on why they have received a specific ad. To obtain such explanations, users need to click on the ``Why am I seeing this?'' button that is in the upper right corner of every ad.  These explanations provide some information regarding the parameters set by the advertisers, but not all~\cite{andreou2018investigating}. We also instrument the browser extension to collect these explanations. 

\paragraph{Ethical considerations:}
We only collect information about the ads and clearly state what we collect to the volunteers who install the extension and accept our terms. We do not collect any information about friends list, likes, photos, videos or regular timeline posts.
Furthermore, the code of our extension is open source and it is publicly available~\footref{dataset}.

\subsection{Dissemination and Deployment}

The initial users who installed our browser extension were friends and family. Later, our browser extension was widely adopted after our project was cited by popular national and international news media outlets such as BBC, El País, Finantial Times, and Folha de São Paulo~\footnote{\url{http://www.eleicoes-sem-fake.dcc.ufmg.br/?section=midia}}. 
Additionally, to disseminate the tool, we  wrote and published opinion articles to  disclose ways to exploit online systems to influence elections and what we can do about them~\cite{fabriciofolha,nyt2018benevenuto}. These risks were then exposed by our team in the Brazilian senate and in multiple national TV shows\footnote{\url{https://www.youtube.com/watch?v=eGScrdi5hhU&t=3450s}.}.




\subsection{Data Collection}

\begin{figure}[t]
\includegraphics[width=0.6\columnwidth]{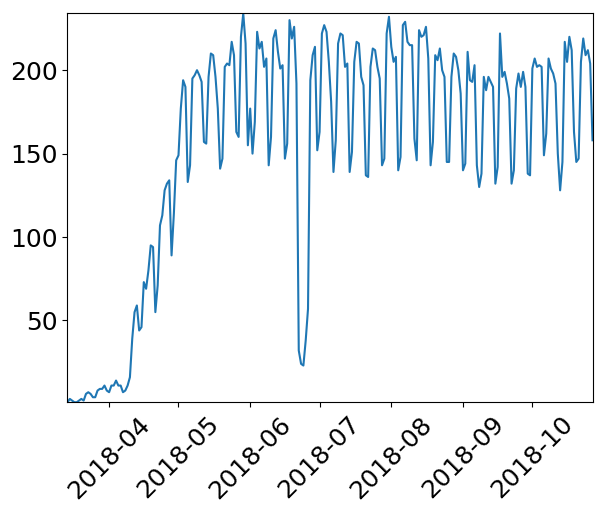}
\caption{Number of daily active users.}
\label{fig:user_by_day}
\end{figure}

We used \adanalystbr to monitor ads from March 14, 2018 to October 28, 2018. This period covers the electoral period, including the two voting rounds. Overall, more than 2,000 users volunteered to install our browser extension and share the ads they received while navigating on Facebook with our project. We noted, however, that many users only installed the browser extension but do not used it. Nevertheless, a total of 715 users actively used our tool in this period.

Figure~\ref{fig:user_by_day} shows the number of active users per day. We consider a user as active in a day if they received at least one Facebook ad. The number of daily users increased rapidly when several news-media outlets published articles about our system and stayed relatively stable afterward. The sudden decrease of active users in mid-June can be attributed to a change Facebook makes in the way it tags ads, resulting in the loss of ads for some days. We also notice that user activity on weekends decreases, which might indicate that some users have installed our plugin in their computer at work.

Out of the 715 users, 682 are from Brazil. We inferred this targeting information by parsing the data from the "\emph{Why am I seeing this?}" explanations that were collected by our extension~\cite{andreou2018investigating}. 
We collected in total \textbf{239k unique ads} sent by \textbf{40k advertisers}. Each ad is identified by an unique \emph{id} provided by Facebook. 
Out of the 239k unique ads, 166k were sent during the pre-electoral period (March 2018 to August 15, 2018) and 74k were sent during the electoral period (August 16, 2018 to October 28, 2018). For each ad we have information about the advertiser, the text of the ad, the text in the political ad disclaimer, the image (when available) and the landing URL. We refer to this dataset as the \emph{\adanalystbr dataset}.

 Official political ads are different than regular ads. Besides the ``Sponsored'' tag, the disclaimer also contains the tag ``Electoral Propaganda''. Because of that, our extension did not collect these ads, i.e., the dataset collected with our browser extension does not contain any official political ad.

\subsection{Facebook Ad Library Dataset}

Facebook provides an Ad Library that includes all ads that were declared as containing political content by the advertisers~\footnote{\url{https://www.facebook.com/politicalcontentads}}. The Ad Library offers a search engine where given a keyword, the engine returns matching advertisers and their ads. Facebook lists the top 30 results for the query and provides a token for the next page.  We implemented a crawler to perform automatic searches and scrape the results provided this application. 
In August 2018, the search engine allowed to query empty strings and was returning a long list of ads sorted by the publish time. We collected all the returned results until the last page. Unfortunately, the Facebook Ad Library changed a lot during the past 2 years, which might make this collection method not possible anymore~\footnote{The task of building and maintaining this crawler is not straightforward.  Facebook constantly updates the URL parameters of search queries and implements countermeasures to detect and block crawlers.}. 

We tried to collect close to all ads declared as political. To do that, we repeated the task of querying for an empty string periodically during the period of Aug 2018 to December 2019, a total of 11 complete crawls.  Each additional crawl increased our dataset by an average of 2.63\%. 
Additionally, to measure how many ads we missed, we performed a hundred searches with distinct random words selected from a Wikipedia dataset~\cite{wikipedia_dataset}. This experiment revealed that our dataset increased, in total, by 2.59\%, or by an average of .0259\% for each search.
While we believe we are close to have the entire dataset of political ads, there is not a systematic way to assess the amount we of ads we did not gathered. As also observed by others, the Facebook Ad Library is unstable and there are ads that appear and ads that disappear each day which makes it impossible to know for sure how many ads we miss~\cite{mozillaadlibraryreport}.

For each ad, Facebook returns the advertiser's name, the advertiser's Brazilian tax id, the ad text, the ad media (video or images), how much money was spent on the ad and information about the age, the gender and the location of the users reached.  We filter out ads that did not reach people in Brazil and are not marked as being about political, election and social issues.\footnote{The new version of the Ad Library also returns active campaigns of advertisers that can include ads that are not political. We exclude these ads from the dataset as well.} 
Our curated dataset contains \textbf{100,778 unique ads} from \textbf{5,292 advertisers} during the electoral period and all ads gathered are specific to Brazil.
We refer to this dataset as the \emph{\Fbdata dataset}.

\section{Detecting political ads} \label{sec:methodology}

There is no consensus on what is a political ad. Different platforms have different definitions for what they consider as political ads~\cite{twitter_ban_political_ad,google_pol_ads}, while at the same time political scholars and regulators are debating about what would be a good definition~\cite{pda}. 

Our goal in this work is not to provide the best definition for a political ad, but to operationalize one that is able to identify ads that are similar to self-declared political ads.  More specifically, our approach consists on  investigating to which extent we can build machine-learning algorithms that are able to accurately identify similar ads from those available in the Facebook political Ad Library. 




In practice, our approach has three key steps. First, we need to create a gold standard collection of ads used to test our method. Then, we select a supervised classifier that receives as input an ad and outputs if it is political , i.e., of the same nature as the ones registered as such on Facebook, or non-political. Finally, we evaluate this classifier over different sets of testing data.

\subsection{Gold standard collection}\label{sec:labeling-ads}

To train and test our machine learning models we need a labeled set of political and non-political ads. Next, we describe our assumptions and how we create a set of political and non-political ads. \\

\noindent
\textbf{Political ads}: As political ads, we choose a uniformly distributed random sample of $10,000$ ads from the \Fbdata dataset.
As these ads were made by political candidates or political parties as part of their official political campaigns, they are good representative instances of political ads for training our machine learning-based methods. 
Our rationale is that, independent of the definition of a political ad, our data-driven approach  might be able to properly recognize an ad that is similar to those self-declared political ads made by real political agents. \\

\noindent
\textbf{Non-political ads}:
There are no available existing labeled datasets for non-political ads. 
Thus, we need to label such ads ourselves. We selected a uniformly distributed random sample of $10,000$ ads from the \adanalystbr dataset. Although most of the ads in this dataset are not about politics, there are a few political ads among them. So, we asked three independent volunteers\footnote{The volunteers were students in our lab.}  to label the ads as political or non-political. We instructed the volunteers to consider as political ads the ads declared on Facebook as \emph{Political and Issue ads}, as well as ads that correspond to the definition proposed by Oliveira \textit{et al.}~\cite{oliveira2018}:
\newtheorem{p_def}{Definition}
\begin{p_def}
\label{def1}
A \textbf{political advertisement} in Facebook is a sponsored message posted on a Facebook page whose content expresses subjects related to state, politics, governance, and justice. Specifically, such messages may cover one or more of the following topics: political campaign; human rights; political activism; political news; federal programs projects and laws; politician public agenda; judicial decisions; public expenditures and crimes against public administration.
\end{p_def}


We evaluated the inter-rater reliability among the independent volunteers using the agreement percentage and the Cohen's Kappa coefficient ($\kappa$)~\cite{Oren2011,Landis1977,Savage2015}. This coefficient measures the agreement between two volunteers who each classify a predefined number of ads as one of the two mutually exclusive categories: \emph{political} or \emph{non-political}. The agreement among the three volunteers is consistently very high, 99.7\% for volunteers 1 and 2, with $\kappa = 0.93$, and 99.5\% for volunteers 1 and 3, with $\kappa = 0.88$. For volunteers 2 and 3 the agreement is 99.5\%, with $\kappa = 0.89$. According to Landis and Koch~\cite{Landis1977}, these Kappa scores fall into the range of scores referred to as ``Almost Perfect'' agreement, which is a satisfactory result that validates our \textit{gold standard collection}.
The labeled data contains $233$ political ads and $9,767$ non-political ads. We added the $233$ political ads as part of our political ads dataset. 
Our final \gs collection is a nearly balanced dataset containing \textbf{10,233} political ads and \textbf{9,767} non-political ads.\footnote{This dataset is publicly available to download at \url{https://lig-membres.imag.fr/gogao/political_ads.html.}}

\subsection{Supervised learning algorithms}

We evaluate five classifying techniques to distinguish between political and non-political ads: Logistic Regression, Random Forest, Support Vector Machine with RBF kernel, and Gradient Boosting using Grid Search for selecting best hyperparameters during the train and Word Embedding for feature extraction, and Naive Bayes using Hashing Vectorizer for feature extraction.

\begin{figure*}[t]
\centering
\includegraphics[width=0.8\linewidth]{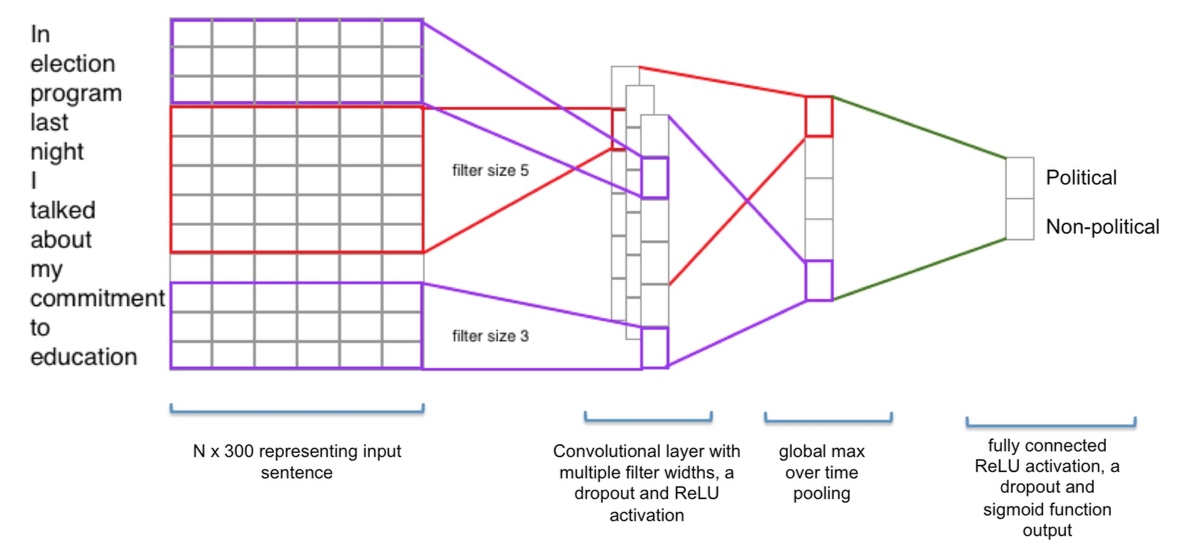}
\caption{Convolutional Neural Network architecture for political ads classification. 
}
\label{fig:cnn_model}
\end{figure*}

We also implemented a Convolution Neural Network (CNN) architecture recently proposed for a similar task: identifying political tweets posted by politicians~\cite{oliveira2018}. Figure~\ref{fig:cnn_model} presents an overview of our implementation.  We represent each word of a Facebook ad as a dense vector retrieved from Word2Vec C-BoW with 300 dimensions, pre-trained over a large Portuguese data set, which is able to produce an embedding matrix for a vocabulary of 1.3 trillion terms~\cite{Mikolov2013,Hartmann2017}. Thus, the input layer is a matrix $n \times 300$, where $n$ is the number of words in a particular ad and $300$ is the vector representation of each word of this message. Subsequently, there is a 25\% rate dropout regularization layer connected to a convolutional layer with $120$ different filters and sizes (3,4,5), activated by a ReLU function. Then, the output of the previous layer is connected to a global max-polling layer, whose output is in turn, fully connected to a ReLU activation and to another 25\% rate dropout layer. Finally, the last dense layer is a single neuron with a sigmoid activation function that outputs $1$ if the message is \textit{political}, and $0$ if \textit{non-political}. We optimized the neural network by means of cross entropy loss function using the RMSProp optimization algorithm~\cite{Duchi2011}.

\subsection{Evaluation}

We use 10-fold cross validation to train and test the six classifiers. We partition the \gs dataset in ten random sub-samples, out of which nine are used as training data, and the remaining one is used for testing the classifier.  The results reported are averages of the 10 runs along with their 90\%-confidence intervals. 

To evaluate the classifiers we considered three classic metrics: accuracy, Macro-F1, and the area under the ROC curve (AUC). Accuracy considers equally important the correct classification of each ad, independently of the class, and basically measures the capability of the method to predict the correct output. As the \gs dataset is nearly balanced, the results for accuracy are meaningful. The resulting AUC is the probability that a model will rank a randomly chosen political ad higher (e.g. more political) than a randomly chosen ad. The AUC is especially relevant for political ad detection since the decision threshold can be used to control the trade-off between true and false positive rates. Finally, Macro-F1 values are computed by first calculating F1 values (a metric that captures both precision and recall for a 0.5 threshold) for each class in isolation, then averaging over all classes. This way, Macro-F1 considers equally important the effectiveness in each class, independently of the relative size of the class. Besides that, the Macro-F1 score is the harmonic mean of the precision and recall.

Table~\ref{tab:results_experiments} presents the accuracy, AUC and Macro-F1 score for the six classifiers we consider. The accuracy of CNN, SVM, Logistic Regression and Naive Bayes is 94\% in a nearly balanced dataset for detecting both political and non-political ads. 



While our classifiers achieve an impressive accuracy in a nearly balanced dataset, in the real-world the number of non-political ads is much higher than the number of political ads. Hence, it is important to study the ROC curves of classifiers and what true positive rates (detection of true political ads) they achieve for small false positive rates (false detection of non-political ads as political ads).  
Table~\ref{tab:results_tpr} shows the true positive rate for a false positive rate of 1\% and 3\%. We observe that CNN and Naive Bayes classifiers are the most accurate with true positive rates of 78\% and 85\% for a 1\% false positive rate.

\begin{table}[t]
\centering
\caption{Accuracy of different classifiers.}
\label{tab:results_experiments}
\begin{tabular}{lccc}
\hline
	\textbf{Classifier} &	\textbf{Accuracy}	&\textbf{AUC} & \textbf{Macro-F1}  \\
\hline
\textbf{CNN} 	        & \boldmath$0.94 (\pm 0.01)$	& \boldmath$0.98 (\pm 0.01)$ & \boldmath$0.94 (\pm 0.01)$ \\ 
\textbf{SVM} 			& \boldmath$0.94 (\pm 0.01)$  & \boldmath$0.98 (\pm 0.01)$ & \boldmath$0.94 (\pm 0.01)$ \\
\textbf{L. Regression} 	& \boldmath$0.94 (\pm 0.01)$  & \boldmath$0.98 (\pm 0.01)$ & \boldmath$0.94 (\pm 0.01)$ \\
R. Forest 		& $0.92 (\pm 0.01)$	& $0.97 (\pm 0.01)$ & $0.92 (\pm 0.01)$\\
\textbf{N. Bayes} 		& \boldmath$0.94 (\pm 0.01)$  & \boldmath$0.99 (\pm 0.01)$ & \boldmath$0.94 (\pm 0.01)$ \\
G. Boosting 	& $0.92 (\pm 0.01)$	& $0.97 (\pm 0.02)$ & $0.92 (\pm 0.01)$ \\

\hline
\end{tabular}
\end{table}

\begin{table}[t]
\centering
\caption{True Positive Rate (TPR) for 1.0\% and 3.0\% False Positive Rate (FPR).}
\label{tab:results_tpr}
\begin{tabular}{lcc}
\hline
	\textbf{Classifier} &	\textbf{FPR = $1\%$}	&\textbf{FPR $3\%$}  \\
\hline
\textbf{CNN} 	        & \boldmath$78\%$	& \boldmath$90\%$ \\ 
SVM 			& $57\%$  & $85\%$ \\
L. Regression 	& $58\%$  & $85\%$  \\
R. Forest 		& $62\%$	& $83\%$ \\
\textbf{N. Bayes} 		& \boldmath$85\%$  & \boldmath$95\%$ \\
G. Boosting 	& $57\%$	& $79\%$ \\

\hline
\end{tabular}
\end{table}

\section{Analysis of political ads}
\label{sec:results}

In this section we investigate how many political ads our classifier
identifies in the \adanalystbr dataset. 
From \adanalystbr we removed
the ads that are part of the \gs dataset and were used for training the models, as well as the ads that we collected outside the election period (March 14, 2018 to August 15, 2018). We also removed  ads that were not in the Portuguese language, resulting in a set of 58,235 unique ads.
Finally, some advertisers create sets of ads that have the same text but with different images. In order to not over-represent them we only kept one ad for each set of ads with multiple images and the same caption. Hence, this leads us to a set of 38,110 ads. Remember that this dataset does not contain any official political ad as our browser extensions did not collect these ads (see Section~\ref{sec:dataset}).

\begin{figure}[t]
\includegraphics[width=0.9\columnwidth]{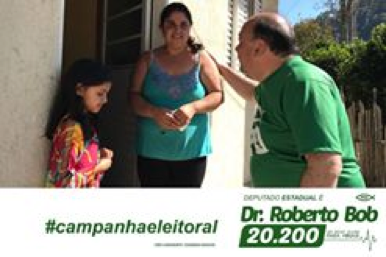}
\caption{Ad text posted together with this ad image translated to English: \normalfont{I am meeting friends, making new friends and bringing my message as a state deputy candidate to all residents of Marmelópolis (MG). Let's go together! A NEW LOOK FOR MINDS Political Advertising: CNPJ 31.208.669/001-05 \#Marmelópolis \#ANewLooking ForMines \#DrRobertoBob20200.}}
\label{fig:ad_sample2}
\end{figure}

We use the CNN model to assign to each ad a probability of being political.
In practice there is a significantly higher number of non-political
ads then political ads (i.e., we have an unbalanced dataset scenario). To limit the number of false positives (i.e., ads that are misclassified as political by our model but are actually non-political) we choose a threshold for declaring an ad as political that corresponds to a 1\% false positive rate (instead of choosing the typical 0.5 threshold for the probability of being political, which corresponds to a false positive rate of 8\% and true positive rate of 96\%). Using a threshold of 0.97, our CNN model classifies 835 ads as political out of the 38,110 ads we tested -- 2\% of the ads are political. The 835 ads were posted by 577 advertisers. 
We name the set of political ads we detect as the \polads. Figure~\ref{fig:ad_sample2} shows an example of such ad. We see that the ad mentions the name of a candidate and his identification number during the election.

To be sure the ads we detect are indeed about political issues we picked a random sample of 300 ads that were targeted by 251 advertisers and we manually check them. We found 19 ads that are not political and 2 ads we are not sure. The ads that were wrongly classified as political had captions that could mostly be easily confused by an algorithm; for example an ice-cream shop was presenting their products as mock-candidates in the election. 

This results suggest that at least 2\% of the ads in our dataset contain political content and are not part of the Facebook Ad Library. Note that our threshold for declaring political ads was very high and there might be many other political ads we do not detect due to this high threshold.  
This means that although a law exists to enforce the disclosure and registration of political ads on Facebook during the elections, there is still a considerable amount of political ads being broadcast that are not disclosed properly.

\vspace{2mm}
\noindent \textbf{Compliance with the Brazilian election law:} 

\noindent \textit{Disclosure:} We observed that some advertisers, even if they do not declare their ad as political on Facebook, they mention in the text of their ad the keywords ``Propaganda Electoral'' or ``Propaganda Política'', and their CPF or CNPJ tax id numbers, as required by the Brazilian law. We identified 90 such ads coming from 53 advertisers in \polads. 
While, these advertisers comply with the Brazilian election law, their ads are not declared to Facebook and, hence, are not  part of the Facebook Ad Library, thus evading future scrutiny. Note that, we have not checked if any of these advertisers declared their spending to the electoral court, to verify if the whole process is compliant.  


\vspace{2mm}
\noindent \textit{Type of advertiser:} 
Only political parties and politicians were allowed by law in Brazil to launch political ads during the electoral period, which is the period considered in our analysis. 
We manually checked who posted each ad from our sample of 300 ads. Out of the 232 advertisers who posted clear political ads, we detected 15 advertisers that are news organizations, and 28 advertisers that were neither news organizations, nor political parties or politicians. 

Many of these advertisers --especially news organizations-- frequently covered news, debate events, interviews e.t.c., however, we  also identified 21 ads (from 12 advertisers) whose message was directly related to politics, often clearly advocating political agendas. For example, an NGO named \emph{Sou Da Paz} was advocating through 9 ads for several agendas related to reducing gun violence, crime rates e.t.c., while a community named \emph{Esquerda Marxista} placed 2 ads with heavy ideological undertone, one of which directly criticized Jair Bolsonaro. Table~\ref{tab:ads_sample} shows a (translated) ad for each of the two advertisers (AD5 and AD6, respectively).




\vspace{2mm}
\noindent \textit{Representation in the Ad Library:} The ads in \polads were not declared as political on Facebook and hence are not part of the Ad Library.  To check whether there might be versions of these ads that are part of the Ad Library, for each ad in \polads we extract the text of the ad and we check if the text matches any of the ads in the \Fbdata dataset. Only 34 of the 835 ads have a corresponding ad in \Fbdata.
This shows that for a small fraction of ads, the advertisers had some ads that were compliant with the Facebook's ToS but also similar versions of the same ad that were not.



\vspace{2mm}
\noindent \textbf{Differences between ads in \polads dataset and the \Fbdata dataset:} 
In Table~\ref{tab:ads_sample} we show (translated) examples of ads from \Fbdata and \polads. 
We do not see a clear distinction among ads of each group as they both contain explicit publicity material for candidates as well as ads related to the elections but not to particular campaigns. This sample of examples illustrates  the importance of having collaborative systems that enable the participation of the society on the process of uncovering (potentially suspicious) political campaign actions.
\begin{table*}[t]
\caption{Sample of ads in \Fbdata and \polads.}
\label{tab:ads_sample}
\begin{small}
\begin{tabular}{ |p{0.5cm} |p{12.5cm}|p{3.5cm}|}
\hline
\# & Ad Text & Ad and Advertiser Info \\
\hline
\multicolumn{3}{c}{\Fbdata} \\
\hline
AD1 & The candidate Romeu Zema, tried once again, to hitch a ride in the campaign of Jair Messias Bolsonaro. This time he didn't make it, because people have already started to discover who he really is. \#eleicoes2018
& Public Figure

fb.com/deputadomarcosmontes
 \\ \hline
 AD2 & 3 days to go! It was an incredible journey to get here receiving affection and support from so many people. Check out the retrospective of my campaign and don't forget: on October 7, vote 2332 for Deputado Federal RJ. I count on you! bit.ly/2P8kkxz
&Public Figure

fb.com/FlavioCavalcantiJunior
\\ \hline
AD3 & It is a great satisfaction to receive the support of Carlos Roberto Osorio. He will not run for reelection, but he demonstrates that he has better security in Rio. Vote 40021 for state deputy - Renan Ferreirinha
& Politician

fb.com/RenanFerreirinha/

\\ \hline 
AD4 & Friends, in this final phase, I need your support! Anyone who has always followed my work knows that I did a lot, but as \#DeputadoEstadual I can do even more. Have you ever chosen11222 help me share it with your friends and family? \#AForçaDoInterior  \#NewPolicy \#WellPolicy \#CleanFile
& Politician

fb.com/fabiomanfrinatobauru/
\\ \hline

\multicolumn{3}{c}{\polads} \\
\hline
AD5 & 
Proposal of Agenda \#SegurançaPúblicaÉSolução: end impunity! Only increasing penalties does not solve impunity: of what use a hundred years worth if the person responsible for the crime is not arrested? The work to change that is to invest in research well made, which identifies criminals, dismantle gangs and arrest those who commit violent crimes such as murder and rape. \#NãoTáTudoBem \#VamosResolver? Meet this and other proposals to improve public security in Brazil: http://bit.ly/ResumoSegurançaPúblicaÉSolução
& NGO

\url{fb.com/institutosoudapaz/}

\\ \hline




AD6 &
The fight against Bolsonaro can only be effective if seen as part of the struggle of the working class against the current system. Come to know the fight of the Communists. \#ElesNão: Bolsonaro and other representatives of capital.
& Community

\url{fb.com/EsquerdaMarxista/}
\\ \hline

AD7 & 
[ELECTIONS 2018 - INTERVIEW WITH BISHOP DAMASCENO] Continuing the interviews of the Luziense Observatory with candidates from Santa Luzia in the 2018 Elections, we talked with the city candidate for Senate, Bishop Damasceno, of the PPL. Check out! \#elections2018 \#interview \#senado \#bispodamasceno \#politics \#TVOL \#santaluzia \#observatorioluziense \#luziar
& News Media

\url{fb.com/observatorioluziense/}
\\ \hline
AD8 & Hi, Hi Dear People! For state representative, vote Luiz Carlos Martins - 11550!
& Politician

\url{fb.com/luizcarlosmartinsoioi/} \\ \hline






\end{tabular}
\end{small}
\end{table*}

\section{Related Work}\label{related-work}

Social networks rise as a new battlefield during elections and the Facebook ad platform was shown to be quite effective in many marketing segments, including political advertising.\footnote{ https://www.wordstream.com/blog/ws/2017/02/28/facebook-advertising-benchmarks} 
We review related efforts along four axes: (i) studies related to sentiment analysis on political content, (ii) studies focused on Facebook ads, (iii) studies on the relationship between social media and the general public, and (iv) studies on the social media influence on election results.

\vspace{2mm}
\noindent{\textbf{Detection of political content and sentiment.}}
Durant \textit{et al.}~\cite{Durant2007} implemented automatic techniques that identify the political sentiment of web blog posts and help bloggers categorize and filter this exploding information source. Bakliwal \textit{et al.}~\cite{bakliwal2013sentiment} also implemented a classifier for sentiment analysis, but they have not detected political content, only sentiments. In contrast, Oliveira \textit{et al.}~\cite{oliveira2018} built a CNN (Convolutional Neural Network) for detecting political tweets from a collection of $2,000$ congressmen tweets labeled as \textit{political} and \textit{non-political}. 

\vspace{2mm}
\noindent{\textbf{Analysis of Facebook ads.}}
Andreou \textit{et al.}~\cite{andreou2018investigating} investigated the level of transparency of Facebook explanations and showed that the Facebook ad explanations are often incomplete and sometimes misleading, while data explanations are often incomplete and vague. In addition, Andreou \textit{et al.}~\cite{Andreou19a} characterized advertises on Facebook. They found that a non-negligible fraction of advertisers belongs to sensitive categories such as news and politics, and that there exist many niche, unverified advertisers whose trustworthiness is difficult to estimate in an automated way. Also, their analyses revealed that a significant amount of advertisers use targeting strategies that can be characterized as invasive or opaque. In a related work, Speicher \textit{et al.}~\cite{speicher-2018-targeted} showed that a malicious advertiser can create highly discriminatory ads without using sensitive attributes.

Recent papers studied how Russian ads were able to affect U.S. citizens. Ribeiro \textit{et al.}~\cite{ribeiro2019@fat} investigated how malicious Russian advertisers were able to run ads with divisive or polarizing
topics (e.g., immigration, race-based policing) at vulnerable subpopulations. Authors have analyzed divisiveness of the ads based on topics that caused different reactions among different social groups. Kim \textit{et al.}~\cite{kim-pol-com2018} used an ad tracking app that enabled them to trace the sponsors/sources of political campaigns and unpack targeting patterns. Their empirical analysis identified ``suspicious'' groups, including foreign entities, and operating divisive issue campaigns on Facebook.
Etudo \textit{et al.}~\cite{etudo2019facebook} also investigated the effects of Russian ads and what is the relation with Black Lives Matter Protests. The study found that Russian ads related to police brutality were issued to coincide with periods of higher unrest.

\vspace{2mm}
\noindent{\textbf{Influence of social media on the general public.}}
Concerning the role of social media on the general public, Wang and Mark \cite{Wang:2017} studied how college students engage with political and social issues on Facebook and found that the impression management and disclosure concerns strongly influence some people to refrain from commenting or sharing content. However, there is evidence that social media can create a public sphere that enables discussions and deliberations~\cite{Maruyama2014}. For instance, Kou et. al. \cite{kou2017one} analyzed the public discourses about Hong Kong’s Umbrella Movement on two distinct social media sites, Facebook and Weibo. They show how people on these two sites reasoned about the many incidents of the movement and developed sometimes similar but other times strikingly different discourses. 

Despite enabling public discourse, social media can leverage some problems such as biased content. For example, Kulshrestha \textit{et al.}~\cite{Kulshrestha:2017} proposed a framework to quantify bias in politics-related queries on Twitter. They found that both the input data and the ranking system contribute significantly to produce varying amounts of bias in the search results, what can have a significant impact on the impression that users form about the different events and politicians. Similarly, Gao \textit{et al.}~\cite{Gao:2018} conducted a controlled experiment to study how stance labels to separate news articles with opposing political ideologies help people explore diverse opinions. Results show that stance labels may intensify selective exposure - a tendency for people to look for agreeable opinions – and make people more vulnerable to polarized opinions and fake news.

These latter two topics are problems analyzed by Che \textit{et al.}~\cite{Che:2018}, who compared and contrasted the ways left- and right-wing news organizations treat the concept of fake news in the context of the highly polarized nature of U.S. news media as well as the evolving and nebulous nature of fake news. They found some key differences. While left-leaning sources discuss specific examples of fake news, the narrative in right-leaning sources focuses on mainstream media as a whole. Moreover, Garimella \textit{et al.}~\cite{Garimella:2018} showed that people who try to bridge the echo chambers, by sharing content with diverse leaning, have to pay a ``price of bipartisanship'', which is a latent phenomenon that effectively stifles mediation between the two sides.

\vspace{2mm}
\noindent{\textbf{Influence of social media on election results.}}
There is already evidence that the action and interaction of voters in social media can influence their inclination to vote or not for a candidate. Maruyama \textit{et al.}~\cite{Maruyama2014} found a relation between Twitter use and the voting choice. They investigated how using a social network while watching a political event could influence the experience of a voter, especially when the user actively participates by posting messages about the event. Pal \textit{et al.}~\cite{pal2018speaking} examined the function and public reception of critical tweeting in online campaigns of four nationalist populist politicians during major national election campaigns. They found that cultural and political differences impact how each politician employs their tactics.
However, politicians are not only the ones who try to influence the elections. Hemphill and Roback \cite{Hemphill2014} examined hundreds of citizen-authored tweets and the development of a categorization scheme to describe common strategies of lobbying on Twitter. Contrary to past research, they found that assumed citizens used Twitter to merely shout out their opinions on issues and utilize a variety of sophisticated techniques to impact political outcomes. Finally, \citet{tumasjan2010predicting} investigated whether Twitter is used as a forum for political deliberation and whether online messages on Twitter validly mirror offline political sentiment. They concluded that the mere number of messages mentioning a party reflects the election result in the German federal election.

Regarding the behavior of politicians in social media, Hwang \cite{Hwang2013} analyzed how Korean young adults evaluate the use of Twitter by South Korean politicians, perceive politicians' credibility, and evaluate politicians who use Twitter. The author concludes that politicians who actively use Twitter are seen as more credible and, as a consequence, are more positively evaluated by young adults.
Still in South Korea, Lee and Shin~\cite{Lee2012} and Lee \textit{et al.}~\cite{Lee2014} designed experiments to investigate how the level of interactivity in politicians' Twitter communication affects the public's cognitive and affective reactions. They found that exposure to high-interactivity Twitter pages induce a stronger sense of direct conversation with the candidate, which, in turn, led to more positive overall evaluations of the candidate and a stronger intention to vote for him. 

Our work is novel in many aspects and provides complementary insights to the discussed studies. 
We focus on detecting political content in advertisement with the aim of monitoring the misuse of Facebook ads platform. Our paper highlights the importance of independent auditing platforms for political ads and our effort provides all the necessary framework to make it feasible.

\section{Concluding Discussion} \label{sec:conclusion}

In this work we presented a system for detecting political ads in Facebook and which we deployed during the Brazilian 2018 elections\footnote{https://www.eleicoessemfake.dcc.ufmg.br/anuncios/}.
Although we have only a small sample of the ads running in Facebook, our approach was able to identify many ads with political content that are not part of the Facebook Ad Library for political ads. 
The main culprit for this situation is that advertisers need to self-declare their political ads as such for them to appear in the Ad Library. It is not clear whether Facebook has any mechanism to enforce compliance. 

The Brazilian election legislation stipulates that ads during the electoral period need to be labeled with the \emph{Propaganda Electoral} tag display the national identification number (CPF for individuals and CNPJ for institutions) of the advertiser. A small fraction of advertisers in our dataset have the right disclaimer stipulated in the Brazilian election law but they did not declare their ads as political to Facebook, hence, they do not appear in the Ad Library.



While the results in our paper are undeniably worrying as there are Brazilian advertisers that did not declare their political ads, there is a positive side to it: we were able to exploit the ads self-declared as political from compliant advertisers to build machine learning-based models that can detect other similar ads coming from advertisers that do not comply with the Facebook's ToS or electoral laws. 

One limitation of our work is that our results refer exclusively to the Brazilian scenario. We only tested machine learning-based models trained on ads in Portuguese and evaluated them only on ads during the 2018 Brazilian elections.  Assessing the accuracy of such techniques on ads in other languages and testing how well these techniques are adapting to future elections remains an open question. 
Because of that, we plan to analyze other elections using the methodology described in this paper and our browser extension continues to be live (\url{https://adanalyst.mpi-sws.org/}). 


We hope our findings and all the real-world experience of deploying a real system along the 2018 Brazilian elections will inform debates around public policies that regulate political advertising on the Internet. 
If a system like ours is implemented on a widespread scale, political campaigns might adopt adversarial strategies that change their marketing strategies in order to exploit our false negative rate. The existence of multiple independent auditing systems would make the monitoring of political ads more robust to attackers. We hope our effort will inspire other initiatives around the world.  Our paper not only highlights the importance of independent auditing platforms for political ads but also provides all necessary framework to make it feasible as our code is open source.

\section{Acknowledgements}

We thank the anonymous reviewers for their helpful comments.
This research was supported in part by Fundação de Amparo à Pesquisa do Estado de Minas Gerais (FAPEMIG), Conselho Nacional de Desenvolvimento Científico e Tecnológico (CNPq), the Data Transparency Lab, and by the French National Research Agency (ANR) through the ANR-17-CE23-0014 grant and MIAI @ Grenoble Alpes, (ANR-19-P3IA-0003).

\bibliographystyle{ACM-Reference-Format}
\bibliography{references}


\begin{thebibliography}{56}


\ifx \showCODEN    \undefined \def \showCODEN     #1{\unskip}     \fi
\ifx \showDOI      \undefined \def \showDOI       #1{#1}\fi
\ifx \showISBNx    \undefined \def \showISBNx     #1{\unskip}     \fi
\ifx \showISBNxiii \undefined \def \showISBNxiii  #1{\unskip}     \fi
\ifx \showISSN     \undefined \def \showISSN      #1{\unskip}     \fi
\ifx \showLCCN     \undefined \def \showLCCN      #1{\unskip}     \fi
\ifx \shownote     \undefined \def \shownote      #1{#1}          \fi
\ifx \showarticletitle \undefined \def \showarticletitle #1{#1}   \fi
\ifx \showURL      \undefined \def \showURL       {\relax}        \fi
\providecommand\bibfield[2]{#2}
\providecommand\bibinfo[2]{#2}
\providecommand\natexlab[1]{#1}
\providecommand\showeprint[2][]{arXiv:#2}

\bibitem[\protect\citeauthoryear{2016 Presidential Campaign Hacking Fast
  Facts}{2016 Presidential Campaign Hacking Fast Facts}{[n.d.]}]%
        {cnnrussian}
2016 Presidential Campaign Hacking Fast Facts
  \bibinfo{year}{[n.d.]}\natexlab{}.
\newblock \bibinfo{title}{2016 Presidential Campaign Hacking Fast Facts}.
\newblock
  \bibinfo{howpublished}{\url{https://edition.cnn.com/2016/12/26/us/2016-presidential-campaign-hacking-fast-facts/index.html}}.
\newblock
\newblock
\shownote{Accessed: 2019-04-04.}


\bibitem[\protect\citeauthoryear{A New Level of Transparency for Ads and
  Pages}{A New Level of Transparency for Ads and Pages}{[n.d.]}]%
        {facebook_ad_archive_launched}
A New Level of Transparency for Ads and Pages
  \bibinfo{year}{[n.d.]}\natexlab{}.
\newblock \bibinfo{title}{A New Level of Transparency for Ads and Pages}.
\newblock
  \bibinfo{howpublished}{\url{https://newsroom.fb.com/news/2018/06/transparency-for-ads-and-pages/}}.
\newblock
\newblock
\shownote{Accessed: 2019-06-27.}


\bibitem[\protect\citeauthoryear{Ad Library Report}{Ad Library
  Report}{[n.d.]}]%
        {mozillaadlibraryreport}
Ad Library Report \bibinfo{year}{[n.d.]}\natexlab{}.
\newblock \bibinfo{title}{Data Collection Log — EU Ad Transparency Report}.
\newblock
  \bibinfo{howpublished}{\url{https://adtransparency.mozilla.org/eu/log/}}.
\newblock
\newblock
\shownote{Accessed: 2020-01-24.}


\bibitem[\protect\citeauthoryear{Andreou, Silva, Benevenuto, Goga, Loiseau, and
  Mislove}{Andreou et~al\mbox{.}}{2019}]%
        {Andreou19a}
\bibfield{author}{\bibinfo{person}{Athanasios Andreou},
  \bibinfo{person}{M\'arcio Silva}, \bibinfo{person}{Fabr\'icio Benevenuto},
  \bibinfo{person}{Oana Goga}, \bibinfo{person}{Patrick Loiseau}, {and}
  \bibinfo{person}{Alan Mislove}.} \bibinfo{year}{2019}\natexlab{}.
\newblock \showarticletitle{Measuring the Facebook Advertising Ecosystem}. In
  \bibinfo{booktitle}{\emph{Proceedings of the Network and Distributed System
  Security Symposium (NDSS)}}.
\newblock


\bibitem[\protect\citeauthoryear{Andreou, Venkatadri, Goga, Gummadi, Loiseau,
  and Mislove}{Andreou et~al\mbox{.}}{2018}]%
        {andreou2018investigating}
\bibfield{author}{\bibinfo{person}{Athanasios Andreou},
  \bibinfo{person}{Giridhari Venkatadri}, \bibinfo{person}{Oana Goga},
  \bibinfo{person}{Krishna~P Gummadi}, \bibinfo{person}{Patrick Loiseau}, {and}
  \bibinfo{person}{Alan Mislove}.} \bibinfo{year}{2018}\natexlab{}.
\newblock \showarticletitle{Investigating ad transparency mechanisms in social
  media: A case study of Facebook's explanations}. In
  \bibinfo{booktitle}{\emph{The Network and Distributed System Security
  Symposium (NDSS)}}.
\newblock


\bibitem[\protect\citeauthoryear{Badawy, Ferrara, and Lerman}{Badawy
  et~al\mbox{.}}{2018}]%
        {badawy2018analyzing}
\bibfield{author}{\bibinfo{person}{Adam Badawy}, \bibinfo{person}{Emilio
  Ferrara}, {and} \bibinfo{person}{Kristina Lerman}.}
  \bibinfo{year}{2018}\natexlab{}.
\newblock \showarticletitle{Analyzing the digital traces of political
  manipulation: The 2016 russian interference twitter campaign}. In
  \bibinfo{booktitle}{\emph{Proceedings of IEEE/ACM International Conference on
  Advances in Social Networks Analysis and Mining (ASONAM)}}.
\newblock


\bibitem[\protect\citeauthoryear{Bakliwal, Foster, van~der Puil, O'Brien,
  Tounsi, and Hughes}{Bakliwal et~al\mbox{.}}{2013}]%
        {bakliwal2013sentiment}
\bibfield{author}{\bibinfo{person}{Akshat Bakliwal}, \bibinfo{person}{Jennifer
  Foster}, \bibinfo{person}{Jennifer van~der Puil}, \bibinfo{person}{Ron
  O'Brien}, \bibinfo{person}{Lamia Tounsi}, {and} \bibinfo{person}{Mark
  Hughes}.} \bibinfo{year}{2013}\natexlab{}.
\newblock \showarticletitle{Sentiment analysis of political tweets: Towards an
  accurate classifier}. In \bibinfo{booktitle}{\emph{Proceedings of the
  Workshop on Language Analysis for Social Media (LASM)}}.
\newblock


\bibitem[\protect\citeauthoryear{Caba{\~n}as, Cuevas, and Cuevas}{Caba{\~n}as
  et~al\mbox{.}}{2018}]%
        {cabanas2018unveiling}
\bibfield{author}{\bibinfo{person}{Jos{\'e}~Gonz{\'a}lez Caba{\~n}as},
  \bibinfo{person}{{\'A}ngel Cuevas}, {and} \bibinfo{person}{Rub{\'e}n
  Cuevas}.} \bibinfo{year}{2018}\natexlab{}.
\newblock \showarticletitle{Unveiling and Quantifying Facebook Exploitation of
  Sensitive Personal Data for Advertising Purposes}. In
  \bibinfo{booktitle}{\emph{Proceedings of the $\{$USENIX$\}$ Security
  Symposium ($\{$USENIX$\}$ Security 18)}}.
\newblock


\bibitem[\protect\citeauthoryear{Campos, Maranhão, and Benevenuto}{Campos
  et~al\mbox{.}}{2018}]%
        {fabriciofolha}
\bibfield{author}{\bibinfo{person}{Ricardo~R. Campos}, \bibinfo{person}{Juliano
  Maranhão}, {and} \bibinfo{person}{Fabrício Benevenuto}.}
  \bibinfo{year}{2018}\natexlab{}.
\newblock \bibinfo{booktitle}{\emph{Fake news and the chronicle of slush fund
  announced}}.
\newblock
\urldef\tempurl%
\url{https://www1.folha.uol.com.br/opiniao/2018/04/ricardo-r-campos-juliano-maranhao-e-fabricio-benevenuto-fake-news-e-a-cronica-do-caixa-2-anunciado.shtml}
\showURL{%
\tempurl}


\bibitem[\protect\citeauthoryear{Che, Metaxa-Kakavouli, and Hancock}{Che
  et~al\mbox{.}}{2018}]%
        {Che:2018}
\bibfield{author}{\bibinfo{person}{Xunru Che}, \bibinfo{person}{Dana\"{e}
  Metaxa-Kakavouli}, {and} \bibinfo{person}{Jeffrey~T. Hancock}.}
  \bibinfo{year}{2018}\natexlab{}.
\newblock \showarticletitle{Fake News in the News: An Analysis of Partisan
  Coverage of the Fake News Phenomenon}. In \bibinfo{booktitle}{\emph{Companion
  of the ACM Conference on Computer Supported Cooperative Work and Social
  Computing (CSCW)}}.
\newblock


\bibitem[\protect\citeauthoryear{Duchi, Hazan, and Singer}{Duchi
  et~al\mbox{.}}{2011}]%
        {Duchi2011}
\bibfield{author}{\bibinfo{person}{John Duchi}, \bibinfo{person}{Elad Hazan},
  {and} \bibinfo{person}{Yoram Singer}.} \bibinfo{year}{2011}\natexlab{}.
\newblock \showarticletitle{{Adaptive Subgradient Methods for Online Learning
  and Stochastic Optimization}}.
\newblock \bibinfo{journal}{\emph{Journal of Machine Learning Research}}
  \bibinfo{volume}{12} (\bibinfo{year}{2011}), \bibinfo{pages}{2121--2159}.
\newblock
\showISBNx{9780982252925}
\showISSN{15324435}
\urldef\tempurl%
\url{https://doi.org/10.1109/CDC.2012.6426698}
\showDOI{\tempurl}
\showeprint[arxiv]{arXiv:1103.4296v1}


\bibitem[\protect\citeauthoryear{Durant and Smith}{Durant and Smith}{2007}]%
        {Durant2007}
\bibfield{author}{\bibinfo{person}{Kathleen~T. Durant} {and}
  \bibinfo{person}{Michael~D. Smith}.} \bibinfo{year}{2007}\natexlab{}.
\newblock \showarticletitle{Predicting the Political Sentiment of Web Log Posts
  Using Supervised Machine Learning Techniques Coupled with Feature Selection}.
  In \bibinfo{booktitle}{\emph{Advances in Web Mining and Web Usage Analysis}}.
\newblock


\bibitem[\protect\citeauthoryear{Etudo, Yoon, and Yaraghi}{Etudo
  et~al\mbox{.}}{2019}]%
        {etudo2019facebook}
\bibfield{author}{\bibinfo{person}{Ugo Etudo}, \bibinfo{person}{Victoria~Y
  Yoon}, {and} \bibinfo{person}{Niam Yaraghi}.}
  \bibinfo{year}{2019}\natexlab{}.
\newblock \showarticletitle{From Facebook to the Streets: Russian Troll Ads and
  Black Lives Matter Protests}. In \bibinfo{booktitle}{\emph{Proceedings of the
  52nd Hawaii International Conference on System Sciences}}.
\newblock


\bibitem[\protect\citeauthoryear{Facebook Ad Archive}{Facebook Ad
  Archive}{[n.d.]}]%
        {facebook_ad_archive}
Facebook Ad Archive \bibinfo{year}{[n.d.]}\natexlab{}.
\newblock \bibinfo{title}{Facebook Ad Archive Plataform}.
\newblock
  \bibinfo{howpublished}{\url{https://www.facebook.com/politicalcontentads}}.
\newblock
\newblock
\shownote{Accessed: 2019-10-14.}


\bibitem[\protect\citeauthoryear{Facebook Ads related to politics}{Facebook Ads
  related to politics}{[n.d.]}]%
        {facebook_political_content}
Facebook Ads related to politics \bibinfo{year}{[n.d.]}\natexlab{}.
\newblock \bibinfo{title}{Facebook: Ads Related to Politics or Issues of
  National Importance.}
\newblock
  \bibinfo{howpublished}{\url{https://www.facebook.com/policies/ads/restricted_content/political}}.
\newblock
\newblock
\shownote{Accessed: 2019-04-04.}


\bibitem[\protect\citeauthoryear{Facebook Ads with Political Content}{Facebook
  Ads with Political Content}{[n.d.]}]%
        {facebook_adarchive_lached}
Facebook Ads with Political Content \bibinfo{year}{[n.d.]}\natexlab{}.
\newblock \bibinfo{title}{Shining a Light on Ads With Political Content}.
\newblock
  \bibinfo{howpublished}{\url{https://newsroom.fb.com/news/2018/05/ads-with-political-content/}}.
\newblock
\newblock
\shownote{Accessed: 2019-04-04.}


\bibitem[\protect\citeauthoryear{Facebook For Business}{Facebook For
  Business}{[n.d.]}]%
        {lookalike}
Facebook For Business \bibinfo{year}{[n.d.]}\natexlab{}.
\newblock \bibinfo{title}{Lookalike Audiences}.
\newblock
  \bibinfo{howpublished}{\url{https://www.facebook.com/business/a/lookalike-audiences}}.
\newblock
\newblock
\shownote{Accessed: 2020-01-13.}


\bibitem[\protect\citeauthoryear{Facebook Issue of national
  importance}{Facebook Issue of national importance}{[n.d.]}]%
        {facebook_issue_content}
Facebook Issue of national importance \bibinfo{year}{[n.d.]}\natexlab{}.
\newblock \bibinfo{title}{Facebook: Issues of national importance}.
\newblock
  \bibinfo{howpublished}{\url{https://www.facebook.com/business/help/214754279118974}}.
\newblock
\newblock
\shownote{Accessed: 2019-04-04.}


\bibitem[\protect\citeauthoryear{Filipe N.~Ribeiro and Redmiles}{Filipe
  N.~Ribeiro and Redmiles}{2019}]%
        {ribeiro2019@fat}
\bibfield{author}{\bibinfo{person}{Mahmoudreza Babaei Lucas Henrique Johnnatan
  Messias Fabr{\'{\i}}cio Benevenuto Oana Goga Krishna P.~Gummadi Filipe
  N.~Ribeiro, Koustuv~Saha} {and} \bibinfo{person}{Elissa~M. Redmiles}.}
  \bibinfo{year}{2019}\natexlab{}.
\newblock \showarticletitle{On Microtargeting Socially Divisive Ads: A Case
  Study of Russia-Linked Ad Campaigns on Facebook}. In
  \bibinfo{booktitle}{\emph{Proceedings of the ACM Conference on Fairness,
  Accountability, and Transparency}} \emph{(\bibinfo{series}{FAT*'19})}.
  \bibinfo{address}{Atlanta, USA}.
\newblock


\bibitem[\protect\citeauthoryear{Gao, Xiao, Karahalios, and Fu}{Gao
  et~al\mbox{.}}{2018}]%
        {Gao:2018}
\bibfield{author}{\bibinfo{person}{Mingkun Gao}, \bibinfo{person}{Ziang Xiao},
  \bibinfo{person}{Karrie Karahalios}, {and} \bibinfo{person}{Wai-Tat Fu}.}
  \bibinfo{year}{2018}\natexlab{}.
\newblock \showarticletitle{To Label or Not to Label: The Effect of Stance and
  Credibility Labels on Readers' Selection and Perception of News Articles}.
\newblock \bibinfo{journal}{\emph{Proc. ACM Hum.-Comput. Interact.}}
  \bibinfo{volume}{2}, \bibinfo{number}{CSCW}, Article \bibinfo{articleno}{55}
  (\bibinfo{date}{Nov.} \bibinfo{year}{2018}), \bibinfo{numpages}{16}~pages.
\newblock
\showISSN{2573-0142}
\urldef\tempurl%
\url{https://doi.org/10.1145/3274324}
\showDOI{\tempurl}


\bibitem[\protect\citeauthoryear{Garimella, De~Francisci~Morales, Gionis, and
  Mathioudakis}{Garimella et~al\mbox{.}}{2018}]%
        {Garimella:2018}
\bibfield{author}{\bibinfo{person}{Kiran Garimella}, \bibinfo{person}{Gianmarco
  De~Francisci~Morales}, \bibinfo{person}{Aristides Gionis}, {and}
  \bibinfo{person}{Michael Mathioudakis}.} \bibinfo{year}{2018}\natexlab{}.
\newblock \showarticletitle{Political Discourse on Social Media: Echo Chambers,
  Gatekeepers, and the Price of Bipartisanship}. In
  \bibinfo{booktitle}{\emph{Proceedings of the 2018 World Wide Web Conference}}
  \emph{(\bibinfo{series}{WWW '18})}. \bibinfo{publisher}{International World
  Wide Web Conferences Steering Committee}, \bibinfo{address}{Republic and
  Canton of Geneva, Switzerland}, \bibinfo{pages}{913--922}.
\newblock
\showISBNx{978-1-4503-5639-8}
\urldef\tempurl%
\url{https://doi.org/10.1145/3178876.3186139}
\showDOI{\tempurl}


\bibitem[\protect\citeauthoryear{Google Ads}{Google Ads}{[n.d.]}]%
        {google}
Google Ads \bibinfo{year}{[n.d.]}\natexlab{}.
\newblock \bibinfo{title}{How Google Ads Works?}
\newblock
  \bibinfo{howpublished}{\url{https://ads.google.com/home/how-it-works/}}.
\newblock
\newblock
\shownote{Accessed: 2019-04-04.}


\bibitem[\protect\citeauthoryear{Google Political Content}{Google Political
  Content}{[n.d.]}]%
        {google_pol_ads}
Google Political Content \bibinfo{year}{[n.d.]}\natexlab{}.
\newblock \bibinfo{title}{Advertising Policies Help - Political Content}.
\newblock
  \bibinfo{howpublished}{\url{https://support.google.com/adspolicy/answer/6014595?hl=en}}.
\newblock
\newblock
\shownote{Accessed: 2020-01-23.}


\bibitem[\protect\citeauthoryear{Hartmann, Fonseca, Shulby, Treviso, Rodrigues,
  and Aluisio}{Hartmann et~al\mbox{.}}{2017}]%
        {Hartmann2017}
\bibfield{author}{\bibinfo{person}{Nathan Hartmann}, \bibinfo{person}{Erick
  Fonseca}, \bibinfo{person}{Christopher Shulby}, \bibinfo{person}{Marcos
  Treviso}, \bibinfo{person}{Jessica Rodrigues}, {and} \bibinfo{person}{Sandra
  Aluisio}.} \bibinfo{year}{2017}\natexlab{}.
\newblock \showarticletitle{{Portuguese Word Embeddings: Evaluating on Word
  Analogies and Natural Language Tasks}}.
\newblock
\showeprint[arxiv]{1708.06025}


\bibitem[\protect\citeauthoryear{Hemphill and Roback}{Hemphill and
  Roback}{2014}]%
        {Hemphill2014}
\bibfield{author}{\bibinfo{person}{Libby Hemphill} {and}
  \bibinfo{person}{Andrew~J. Roback}.} \bibinfo{year}{2014}\natexlab{}.
\newblock \showarticletitle{Tweet Acts: How Constituents Lobby Congress via
  Twitter}. In \bibinfo{booktitle}{\emph{Proceedings of the 17th ACM Conference
  on Computer Supported Cooperative Work \&\#38; Social Computing}}
  \emph{(\bibinfo{series}{CSCW '14})}. \bibinfo{publisher}{ACM},
  \bibinfo{address}{New York, NY, USA}, \bibinfo{pages}{1200--1210}.
\newblock
\showISBNx{978-1-4503-2540-0}
\urldef\tempurl%
\url{https://doi.org/10.1145/2531602.2531735}
\showDOI{\tempurl}


\bibitem[\protect\citeauthoryear{Hwang}{Hwang}{2013}]%
        {Hwang2013}
\bibfield{author}{\bibinfo{person}{Sungwook Hwang}.}
  \bibinfo{year}{2013}\natexlab{}.
\newblock \showarticletitle{{The Effect of Twitter Use on Politicians'
  Credibility and Attitudes toward Politicians}}.
\newblock \bibinfo{journal}{\emph{Journal of Public Relations Research}}
  \bibinfo{volume}{25}, \bibinfo{number}{3} (\bibinfo{year}{2013}),
  \bibinfo{pages}{246--258}.
\newblock
\showISBNx{1062-726X}
\showISSN{1062726X}
\urldef\tempurl%
\url{https://doi.org/10.1080/1062726X.2013.788445}
\showDOI{\tempurl}


\bibitem[\protect\citeauthoryear{Kim, Hsu, Neiman, Kou, Bankston, Kim,
  Heinrich, Baragwanath, and Raskutti}{Kim et~al\mbox{.}}{2018}]%
        {kim-pol-com2018}
\bibfield{author}{\bibinfo{person}{Young~Mie Kim}, \bibinfo{person}{Jordan
  Hsu}, \bibinfo{person}{David Neiman}, \bibinfo{person}{Colin Kou},
  \bibinfo{person}{Levi Bankston}, \bibinfo{person}{Soo~Yun Kim},
  \bibinfo{person}{Richard Heinrich}, \bibinfo{person}{Robyn Baragwanath},
  {and} \bibinfo{person}{Garvesh Raskutti}.} \bibinfo{year}{2018}\natexlab{}.
\newblock \showarticletitle{The stealth media? Groups and targets behind
  divisive issue campaigns on Facebook}.
\newblock \bibinfo{journal}{\emph{Political Communication}}
  \bibinfo{volume}{35}, \bibinfo{number}{4} (\bibinfo{year}{2018}),
  \bibinfo{pages}{515--541}.
\newblock


\bibitem[\protect\citeauthoryear{Kou, Kow, Gui, and Cheng}{Kou
  et~al\mbox{.}}{2017}]%
        {kou2017one}
\bibfield{author}{\bibinfo{person}{Yubo Kou}, \bibinfo{person}{Yong~Ming Kow},
  \bibinfo{person}{Xinning Gui}, {and} \bibinfo{person}{Waikuen Cheng}.}
  \bibinfo{year}{2017}\natexlab{}.
\newblock \showarticletitle{One social movement, two social media sites: A
  comparative study of public discourses}.
\newblock \bibinfo{journal}{\emph{Computer Supported Cooperative Work (CSCW)}}
  \bibinfo{volume}{26}, \bibinfo{number}{4-6} (\bibinfo{year}{2017}),
  \bibinfo{pages}{807--836}.
\newblock


\bibitem[\protect\citeauthoryear{Kulshrestha, Eslami, Messias, Zafar, Ghosh,
  Gummadi, and Karahalios}{Kulshrestha et~al\mbox{.}}{2017}]%
        {Kulshrestha:2017}
\bibfield{author}{\bibinfo{person}{Juhi Kulshrestha},
  \bibinfo{person}{Motahhare Eslami}, \bibinfo{person}{Johnnatan Messias},
  \bibinfo{person}{Muhammad~Bilal Zafar}, \bibinfo{person}{Saptarshi Ghosh},
  \bibinfo{person}{Krishna~P. Gummadi}, {and} \bibinfo{person}{Karrie
  Karahalios}.} \bibinfo{year}{2017}\natexlab{}.
\newblock \showarticletitle{Quantifying Search Bias: Investigating Sources of
  Bias for Political Searches in Social Media}. In
  \bibinfo{booktitle}{\emph{Proceedings of the 2017 ACM Conference on Computer
  Supported Cooperative Work and Social Computing}}
  \emph{(\bibinfo{series}{CSCW '17})}. \bibinfo{publisher}{ACM},
  \bibinfo{address}{New York, NY, USA}, \bibinfo{pages}{417--432}.
\newblock
\showISBNx{978-1-4503-4335-0}
\urldef\tempurl%
\url{https://doi.org/10.1145/2998181.2998321}
\showDOI{\tempurl}


\bibitem[\protect\citeauthoryear{Landis and Koch}{Landis and Koch}{1977}]%
        {Landis1977}
\bibfield{author}{\bibinfo{person}{J.~Richard Landis} {and}
  \bibinfo{person}{Gary~G. Koch}.} \bibinfo{year}{1977}\natexlab{}.
\newblock \showarticletitle{{The Measurement of Observer Agreement for
  Categorical Data}}.
\newblock \bibinfo{journal}{\emph{Biometrics}} \bibinfo{volume}{33},
  \bibinfo{number}{1} (\bibinfo{year}{1977}), \bibinfo{pages}{159}.
\newblock
\showISBNx{0006341X}
\showISSN{0006341X}
\urldef\tempurl%
\url{https://doi.org/10.2307/2529310}
\showDOI{\tempurl}
\showeprint[arxiv]{NIHMS150003}


\bibitem[\protect\citeauthoryear{Leathern}{Leathern}{2018}]%
        {tos_change_facebook}
\bibfield{author}{\bibinfo{person}{Rob Leathern}.}
  \bibinfo{year}{2018}\natexlab{}.
\newblock \bibinfo{title}{{Shining a Light on Ads With Political Content}}.
\newblock
\newblock
\urldef\tempurl%
\url{https://about.fb.com/news/2018/05/ads-with-political-content/}
\showURL{%
\tempurl}


\bibitem[\protect\citeauthoryear{Lee and Shin}{Lee and Shin}{2012}]%
        {Lee2012}
\bibfield{author}{\bibinfo{person}{Eun-Ju Lee} {and} \bibinfo{person}{Soo~Yun
  Shin}.} \bibinfo{year}{2012}\natexlab{}.
\newblock \showarticletitle{{Are They Talking to Me? Cognitive and Affective
  Effects of Interactivity in Politicians' Twitter Communication}}.
\newblock \bibinfo{journal}{\emph{Cyberpsychology, Behavior, and Social
  Networking}} \bibinfo{volume}{15}, \bibinfo{number}{10}
  (\bibinfo{year}{2012}), \bibinfo{pages}{515--520}.
\newblock
\showISBNx{21522715}
\showISSN{2152-2715}
\urldef\tempurl%
\url{https://doi.org/10.1089/cyber.2012.0228}
\showDOI{\tempurl}


\bibitem[\protect\citeauthoryear{Lee and Shin}{Lee and Shin}{2014}]%
        {Lee2014}
\bibfield{author}{\bibinfo{person}{Eun~Ju Lee} {and} \bibinfo{person}{Soo~Yun
  Shin}.} \bibinfo{year}{2014}\natexlab{}.
\newblock \showarticletitle{{When the Medium Is the Message: How
  Transportability Moderates the Effects of Politicians' Twitter
  Communication}}.
\newblock \bibinfo{journal}{\emph{Communication Research}}
  \bibinfo{volume}{41}, \bibinfo{number}{8} (\bibinfo{year}{2014}),
  \bibinfo{pages}{1088--1110}.
\newblock
\showISBNx{0093650212466}
\showISSN{15523810}
\urldef\tempurl%
\url{https://doi.org/10.1177/0093650212466407}
\showDOI{\tempurl}


\bibitem[\protect\citeauthoryear{Lima, Reis, Melo, Murai, Araujo, Vikatos, and
  Benevenuto}{Lima et~al\mbox{.}}{2018}]%
        {lima2018@asonam}
\bibfield{author}{\bibinfo{person}{Lucas Lima}, \bibinfo{person}{Julio C.~S.
  Reis}, \bibinfo{person}{Philipe Melo}, \bibinfo{person}{Fabricio Murai},
  \bibinfo{person}{Leandro Araujo}, \bibinfo{person}{Pantelis Vikatos}, {and}
  \bibinfo{person}{Fabricio Benevenuto}.} \bibinfo{year}{2018}\natexlab{}.
\newblock \showarticletitle{Inside the Right-Leaning Echo Chambers:
  Characterizing Gab, an Unmoderated Social System}. In
  \bibinfo{booktitle}{\emph{Proceedings of the 2018 IEEE/ACM International
  Conference on Advances in Social Networks Analysis and Mining}}
  \emph{(\bibinfo{series}{ASONAM'18})}.
\newblock


\bibitem[\protect\citeauthoryear{lokkalike}{lokkalike}{[n.d.]}]%
        {cnnAdLibraryDontWork}
lokkalike \bibinfo{year}{[n.d.]}\natexlab{}.
\newblock \bibinfo{title}{Ad Tool Facebook Built to Fight Disinformation
  Doesn’t Work as Advertised}.
\newblock
  \bibinfo{howpublished}{\url{https://www.nytimes.com/2019/07/25/technology/facebook-ad-library.html}}.
\newblock
\newblock
\shownote{Accessed: 2020-01-24.}


\bibitem[\protect\citeauthoryear{Maruyama, Robertson, Douglas, Semaan, Faucett,
  and Program}{Maruyama et~al\mbox{.}}{2014}]%
        {Maruyama2014}
\bibfield{author}{\bibinfo{person}{Misa Maruyama}, \bibinfo{person}{Scott~P
  Robertson}, \bibinfo{person}{Sara Douglas}, \bibinfo{person}{Bryan Semaan},
  \bibinfo{person}{Heather Faucett}, {and}
  \bibinfo{person}{Information~Sciences Program}.}
  \bibinfo{year}{2014}\natexlab{}.
\newblock \showarticletitle{{Hybrid Media Consumption : How Tweeting During a
  Televised Political Debate Influences the Vote}}.
\newblock \bibinfo{journal}{\emph{Proceedings of the 17th ACM conference on
  Computer supported cooperative work {\&} social computing - CSCW '14}}
  (\bibinfo{year}{2014}), \bibinfo{pages}{1422--1432}.
\newblock
\showISBNx{9781450325400}
\urldef\tempurl%
\url{https://doi.org/10.1145/2531602.2531719}
\showDOI{\tempurl}


\bibitem[\protect\citeauthoryear{Mikolov, Chen, Corrado, and Dean}{Mikolov
  et~al\mbox{.}}{2013}]%
        {Mikolov2013}
\bibfield{author}{\bibinfo{person}{Tomas Mikolov}, \bibinfo{person}{Kai Chen},
  \bibinfo{person}{Greg Corrado}, {and} \bibinfo{person}{Jeffrey Dean}.}
  \bibinfo{year}{2013}\natexlab{}.
\newblock \showarticletitle{{Efficient Estimation of Word Representations in
  Vector Space}}.
\newblock \bibinfo{journal}{\emph{Arxiv}} (\bibinfo{year}{2013}),
  \bibinfo{pages}{1--12}.
\newblock
\showISBNx{1532-4435}
\showISSN{15324435}
\urldef\tempurl%
\url{https://doi.org/10.1162/153244303322533223}
\showDOI{\tempurl}
\showeprint[arxiv]{arXiv:1301.3781v3}


\bibitem[\protect\citeauthoryear{Mozilla Blog}{Mozilla Blog}{[n.d.]}]%
        {efective_ad_archive}
Mozilla Blog \bibinfo{year}{[n.d.]}\natexlab{}.
\newblock \bibinfo{title}{Facebook and Google: This is What an Effective Ad
  Archive API Looks Like}.
\newblock
  \bibinfo{howpublished}{\url{https://blog.mozilla.org/blog/2019/03/27/facebook-and-google-this-is-what-an-effective-ad-archive-api-looks-like/}}.
\newblock
\newblock
\shownote{Accessed: 2019-06-26.}


\bibitem[\protect\citeauthoryear{Nytimes}{Nytimes}{[n.d.]}]%
        {russian-ads}
Nytimes \bibinfo{year}{[n.d.]}\natexlab{}.
\newblock \bibinfo{title}{These Are the Ads Russia Bought on Facebook in 2016}.
\newblock
  \bibinfo{howpublished}{\url{https://www.nytimes.com/2017/11/01/us/politics/russia-2016-election-facebook.html}}.
\newblock
\newblock
\shownote{Accessed: 2019-04-04.}


\bibitem[\protect\citeauthoryear{Oliveira, de~Melo, Amaral, and Pinho}{Oliveira
  et~al\mbox{.}}{2018}]%
        {oliveira2018}
\bibfield{author}{\bibinfo{person}{Lucas~S. Oliveira},
  \bibinfo{person}{Pedro~Vaz de Melo}, \bibinfo{person}{Marcelo Amaral}, {and}
  \bibinfo{person}{Jos{\'e}~Ant{\^ o}nio Pinho}.}
  \bibinfo{year}{2018}\natexlab{}.
\newblock \showarticletitle{When Politicians Talk About Politics: Identifying
  Political Tweets of Brazilian Congressmen}.
\newblock \bibinfo{journal}{\emph{International AAAI Conference on Web and
  Social Media}} (\bibinfo{year}{2018}).
\newblock


\bibitem[\protect\citeauthoryear{Oren and Gilbert}{Oren and Gilbert}{2011}]%
        {Oren2011}
\bibfield{author}{\bibinfo{person}{Michael~A Oren} {and}
  \bibinfo{person}{Stephen~B Gilbert}.} \bibinfo{year}{2011}\natexlab{}.
\newblock \showarticletitle{{Framework for measuring social affinity for CSCW
  software}}.
\newblock \bibinfo{journal}{\emph{CHI EA '11: CHI '11 Extended Abstracts on
  Human Factors in Computing Systems}} (\bibinfo{year}{2011}),
  \bibinfo{pages}{1387--1392}.
\newblock
\showISBNx{9781450302685}
\urldef\tempurl%
\url{https://doi.org/10.1145/1979742.1979779}
\showDOI{\tempurl}


\bibitem[\protect\citeauthoryear{Pal, Thawani, Van Der~Vlugt, Out, Chandra,
  et~al\mbox{.}}{Pal et~al\mbox{.}}{2018}]%
        {pal2018speaking}
\bibfield{author}{\bibinfo{person}{Joyojeet Pal}, \bibinfo{person}{Udit
  Thawani}, \bibinfo{person}{Elmer Van Der~Vlugt}, \bibinfo{person}{Wim Out},
  \bibinfo{person}{Priyank Chandra}, {et~al\mbox{.}}}
  \bibinfo{year}{2018}\natexlab{}.
\newblock \showarticletitle{Speaking their Mind: Populist Style and
  Antagonistic Messaging in the Tweets of Donald Trump, Narendra Modi, Nigel
  Farage, and Geert Wilders}.
\newblock \bibinfo{journal}{\emph{Computer Supported Cooperative Work (CSCW)}}
  \bibinfo{volume}{27}, \bibinfo{number}{3-6} (\bibinfo{year}{2018}),
  \bibinfo{pages}{293--326}.
\newblock


\bibitem[\protect\citeauthoryear{pda}{pda}{[n.d.]}]%
        {pda}
pda \bibinfo{year}{[n.d.]}\natexlab{}.
\newblock \bibinfo{title}{Political Advertising - Regulation Candidates,
  Campaigns, and Lobbyists}.
\newblock
  \bibinfo{howpublished}{\url{https://www.pdc.wa.gov/learn/publications/political-committee-instructions/political-advertising}}.
\newblock
\newblock
\shownote{Accessed: 2020-01-24.}


\bibitem[\protect\citeauthoryear{Political Content on Twitter}{Political
  Content on Twitter}{[n.d.]}]%
        {twitter_ban_political_ad}
Political Content on Twitter \bibinfo{year}{[n.d.]}\natexlab{}.
\newblock \bibinfo{title}{Twitter - Prohibited Content Policies}.
\newblock
  \bibinfo{howpublished}{\url{https://business.twitter.com/en/help/ads-policies/prohibited-content-policies/political-content.html}}.
\newblock
\newblock
\shownote{Accessed: 2020-01-19.}


\bibitem[\protect\citeauthoryear{Propublica Website}{Propublica
  Website}{[n.d.]}]%
        {propublicahome}
Propublica Website \bibinfo{year}{[n.d.]}\natexlab{}.
\newblock \bibinfo{title}{Propublica}.
\newblock
  \bibinfo{howpublished}{\url{https://projects.propublica.org/facebook-ads/}}.
\newblock
\newblock
\shownote{Accessed: 2020-01-23.}


\bibitem[\protect\citeauthoryear{Savage, Monroy-Hernandez, and Acm}{Savage
  et~al\mbox{.}}{2015}]%
        {Savage2015}
\bibfield{author}{\bibinfo{person}{S Savage}, \bibinfo{person}{A
  Monroy-Hernandez}, {and} \bibinfo{person}{Acm}.}
  \bibinfo{year}{2015}\natexlab{}.
\newblock \showarticletitle{{Participatory Militias: An Analysis of an Armed
  Movement's Online Audience}}.
\newblock \bibinfo{journal}{\emph{Proceedings of the 2015 Acm International
  Conference on Computer-Supported Cooperative Work and Social Computing
  (Cscw'15)}} (\bibinfo{year}{2015}), \bibinfo{pages}{724--733}.
\newblock
\showISBNx{9781450329224}
\urldef\tempurl%
\url{https://doi.org/10.1145/2675133.2675295}
\showDOI{\tempurl}
\showeprint[arxiv]{arXiv:1502.02065v1}


\bibitem[\protect\citeauthoryear{Speicher, Ali, Venkatadri, Ribeiro,
  Arvanitakis, Benevenuto, Gummadi, Loiseau, and Mislove}{Speicher
  et~al\mbox{.}}{2018}]%
        {speicher-2018-targeted}
\bibfield{author}{\bibinfo{person}{Till Speicher}, \bibinfo{person}{Muhammad
  Ali}, \bibinfo{person}{Giridhari Venkatadri}, \bibinfo{person}{Filipe~Nunes
  Ribeiro}, \bibinfo{person}{George Arvanitakis}, \bibinfo{person}{Fabricio
  Benevenuto}, \bibinfo{person}{Krishna~P. Gummadi}, \bibinfo{person}{Patrick
  Loiseau}, {and} \bibinfo{person}{Alan Mislove}.}
  \bibinfo{year}{2018}\natexlab{}.
\newblock \showarticletitle{{On the Potential for Discrimination in Online
  Targeted Advertising}}. In \bibinfo{booktitle}{\emph{Proceedings of the
  Conference on Fairness, Accountability, and Transparency (FAT*'18)}}.
\newblock


\bibitem[\protect\citeauthoryear{Tardaguila, Benevenuto, and
  Ortellado}{Tardaguila et~al\mbox{.}}{2018}]%
        {nyt2018benevenuto}
\bibfield{author}{\bibinfo{person}{Cristina Tardaguila},
  \bibinfo{person}{Fabricio Benevenuto}, {and} \bibinfo{person}{Pablo
  Ortellado}.} \bibinfo{year}{2018}\natexlab{}.
\newblock \bibinfo{booktitle}{\emph{Fake News Is Poisoning Brazilian Politics.
  WhatsApp Can Stop It}}.
\newblock
\urldef\tempurl%
\url{https://www.nytimes.com/2018/10/17/opinion/brazil-election-fake-news-whatsapp.html}
\showURL{%
\tempurl}


\bibitem[\protect\citeauthoryear{The regulation of online political
  micro-targeting in Europe}{The regulation of online political micro-targeting
  in Europe}{[n.d.]}]%
        {microtargeting_europe}
The regulation of online political micro-targeting in Europe
  \bibinfo{year}{[n.d.]}\natexlab{}.
\newblock \bibinfo{title}{The regulation of online political micro-targeting in
  Europe}.
\newblock
  \bibinfo{howpublished}{\url{https://pure.uva.nl/ws/files/44117536/Internet_Policy_Review_The_regulation_of_online_political_micro_targeting_in_Europe_2020_01_16.pdf}}.
\newblock
\newblock
\shownote{Accessed: 2019-10-14.}


\bibitem[\protect\citeauthoryear{tospolitical}{tospolitical}{[n.d.]}]%
        {tospolitical}
tospolitical \bibinfo{year}{[n.d.]}\natexlab{}.
\newblock \bibinfo{title}{Facebook ads - Anúncios relacionados a temas
  sociais, eleições ou política}.
\newblock
  \bibinfo{howpublished}{\url{https://www.facebook.com/business/help/1256706951128601}}.
\newblock
\newblock
\shownote{Accessed: 2020-01-24.}


\bibitem[\protect\citeauthoryear{TSE - Electoral Court}{TSE - Electoral
  Court}{[n.d.]}]%
        {electoral_law}
TSE - Electoral Court \bibinfo{year}{[n.d.]}\natexlab{}.
\newblock \bibinfo{title}{Brazilian Law Nº 13.488, Octuber 6, 2017}.
\newblock
  \bibinfo{howpublished}{\url{http://www.justicaeleitoral.jus.br/arquivos/propaganda-eleitoral-na-internet}}.
\newblock
\newblock
\shownote{Accessed: 2019-10-14.}


\bibitem[\protect\citeauthoryear{Tumasjan, Sprenger, Sandner, and
  Welpe}{Tumasjan et~al\mbox{.}}{2010}]%
        {tumasjan2010predicting}
\bibfield{author}{\bibinfo{person}{Andranik Tumasjan},
  \bibinfo{person}{Timm~Oliver Sprenger}, \bibinfo{person}{Philipp~G Sandner},
  {and} \bibinfo{person}{Isabell~M Welpe}.} \bibinfo{year}{2010}\natexlab{}.
\newblock \showarticletitle{Predicting elections with twitter: What 140
  characters reveal about political sentiment.}
\newblock \bibinfo{journal}{\emph{Icwsm}} \bibinfo{volume}{10},
  \bibinfo{number}{1} (\bibinfo{year}{2010}), \bibinfo{pages}{178--185}.
\newblock


\bibitem[\protect\citeauthoryear{Wang and Mark}{Wang and Mark}{2017}]%
        {Wang:2017}
\bibfield{author}{\bibinfo{person}{Yiran Wang} {and} \bibinfo{person}{Gloria
  Mark}.} \bibinfo{year}{2017}\natexlab{}.
\newblock \showarticletitle{Engaging with Political and Social Issues on
  Facebook in College Life}. In \bibinfo{booktitle}{\emph{Proceedings of the
  2017 ACM Conference on Computer Supported Cooperative Work and Social
  Computing}} \emph{(\bibinfo{series}{CSCW '17})}. \bibinfo{publisher}{ACM},
  \bibinfo{address}{New York, NY, USA}, \bibinfo{pages}{433--445}.
\newblock
\showISBNx{978-1-4503-4335-0}
\urldef\tempurl%
\url{https://doi.org/10.1145/2998181.2998295}
\showDOI{\tempurl}


\bibitem[\protect\citeauthoryear{Website Who Target Me}{Website Who Target
  Me}{[n.d.]}]%
        {whotargetsme}
Website Who Target Me \bibinfo{year}{[n.d.]}\natexlab{}.
\newblock \bibinfo{title}{Who Targets Me}.
\newblock \bibinfo{howpublished}{\url{https://whotargets.me/en/}}.
\newblock
\newblock
\shownote{Accessed: 2020-01-23.}


\bibitem[\protect\citeauthoryear{whatispolitical}{whatispolitical}{[n.d.]}]%
        {whatispoliticalfacebook}
whatispolitical \bibinfo{year}{[n.d.]}\natexlab{}.
\newblock \bibinfo{title}{Hard Questions: Why Doesn’t Facebook Just Ban
  Political Ads?}
\newblock
  \bibinfo{howpublished}{\url{https://about.fb.com/news/2018/05/hard-questions-political-ads/}}.
\newblock
\newblock
\shownote{Accessed: 2020-01-24.}


\bibitem[\protect\citeauthoryear{Wikipedia Dump}{Wikipedia Dump}{[n.d.]}]%
        {wikipedia_dataset}
Wikipedia Dump \bibinfo{year}{[n.d.]}\natexlab{}.
\newblock \bibinfo{title}{Wikipedia Portuguese Dataset}.
\newblock
  \bibinfo{howpublished}{\url{https://www.wikidata.org/wiki/Wikidata:Database_download/pt-br}}.
\newblock
\newblock
\shownote{Accessed: 2020-01-13.}


\end{thebibliography}

\end{document}